# Chaos Models in Economics

Sorin Vlad, Paul Pascu and Nicolae Morariu

**Abstract**—The paper discusses the main ideas of the chaos theory and presents mainly the importance of the nonlinearities in the mathematical models. Chaos and order are apparently two opposite terms. The fact that in chaos can be found a certain precise symmetry (Feigenbaum numbers) is even more surprising. As an illustration of the ubiquity of chaos, three models among many other existing models that have chaotic features are presented here: the nonlinear feedback profit model, one model for the simulation of the exchange rate and one application of the chaos theory in the capital markets.

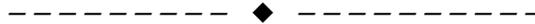

## 1 INTRODUCTION

ONE of the axioms of the modern science asserts that if an accurate description of a physical system can be identified then the possibility of a deeper understanding of the system and the prediction of the system evolution is possible. These assertions are not always correct. For instance, if one applies the laws of motion stated by Newton, then there is possible to predict exactly the orbit of the Moon around the Earth if the influence of other planets is not considered. These predictions were verified and proved to be accurate. If the third planet is included, the mathematical model of the interaction of the two bodies becomes "the three bodies problem", solved by Newton but for a limited set of cases and unsolved for the general case. Today by means of a computer, "the tree bodies problem" can be solved, but one can observe that the prediction of the orbit of the third planet is often impossible.

A large number of real systems have a nonlinear behavior despite the idealized linear behavior used in modeling. The development of a new way of dealing with nonlinear systems is obvious. This "new way of dealing" exists already despite the fact that the study of the nonlinearity is still at the beginning.

Some changes in nonlinear systems can lead to a complex and erratic behavior called chaos. The nonlinearity is one of the conditions needed by a system in order to develop chaos. The term chaos is used to describe the behavior of a system that is aperiodic and apparently random.

S. H. Strogatz defines chaos as an aperiodic long time behavior developed by a deterministic system highly sensitive on initial condition. [1] Behind this apparently random behavior lies the deterministic character determined by the equations describing the system. Most of the systems that are used as examples to explain the concepts of chaos theory are deterministic.

There are two types of chaos: deterministic and nondeterministic. The deterministic chaos represents the chaotic motion of the nonlinear systems whose dynamic laws determines uniquely the evolution of the system's state based on the previous evolution.

The deterministic chaos represents only one particular case of what is called nondeterministic chaos that exhibits a superexponential divergence of the trajectories. In this case the equations describing the evolution of the system are not known. The both ways of chaos manifestations are short-term predictable but long term unpredictable.

The chaos and the concepts related to the dynamics of the systems and the their modeling using differential equations is named the chaos theory and is tightly related with the notion of nonlinearity [4]. The nonlinearity implies the loss of the causality correlation between the perturbation and effect propagated in time. The study of the nonlinearity is named nonlinear dynamics – a captivating domain using a mathematical apparatus still under development.

Despite the fact that the ideas leading to the emergence of the chaos theory existed before longtime, Lorenz (1963) created a mathematical model of the convection currents circulation in atmosphere and observed that when the systems begins with initial conditions slightly changed from the previous ones, the results are completely different. This phenomenon will lie at the basis of a very popular paradigm of chaos named "the butterfly effect", that states that if the flapping of a butterfly slightly modifies the atmospherically conditions in the Amazonian jungle, this fact can have an impact, at the end of a complex cause – effect chain in setting off a tornado in Texas.

The butterfly effect paradigm contains the essence of the phenomenon characterizing the chaos: first, the sensitive dependence on initial conditions and second – the fact that to predict the future state of a chaotic system, the current state need to be known with infinite prediction. The manifestation of chaos can be found everywhere in the real world, for instance: the propagation of the avalanches, epidemics spreading, climate evolution, heart beats, lasers, electronic circuits, etc.

————————————————

- *S. Vlad is with University of Suceava, 13 University Street, Suceava, Romania, 720229.*
- *P. Pascu is with University of Suceava, 13 University Street, Suceava, Romania, 720229.*
- *N. Morariu is with University of Suceava, 13 University Street, Suceava, Romania, 720229.*

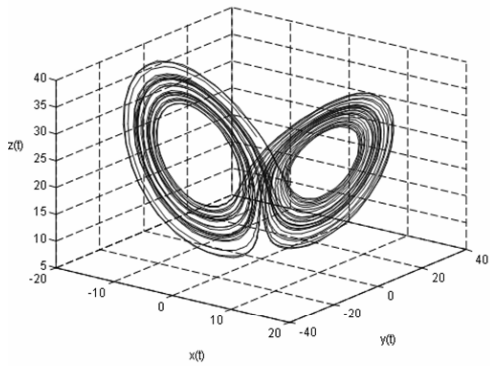

Fig. 1. Lorentz attractor – the symbol of chaos theory.

A legitimate question is that the chaos is the rule or the exception from the rule. Taking into account that most of the systems of the real world are nonlinear (the basic condition for the emergence of chaos), seems that chaos could be one of the not so obvious features of the nature.

The importance of studying chaos is that chaos offers an alternate method that explains the apparently random behavior of the complex systems. The chaos plus the specific mathematical tools is a framework of studying different models from different fields, models that can be reduced to elementary models with known chaotic behavior for some values of the parameters.

The way to chaos begins with the phenomenon of period doubling. The period doubling evolves in 2, 4, 8, 16 and so on periods and the system evolution can abruptly fall into chaotic regime.

In the case of unimodal function there is an interesting symmetry in the parameter values for what the period doubling occurs.

If $A_1$ is the value of the control parameter for what the first period doubling occurs and $A_n$ is the value for what the $n^{th}$ period doubling occurs, then:

$$\delta = \lim_{n \to \infty} \frac{A_n - A_{n-1}}{A_{n+1} - A_n} = 4.66920 \quad (1)$$

where δ is the Feigenbaum number valable for all unimodal functions.[5]

## 2 NONLINEAR MODELS

### 2.1 Chaos in exchange rates

For the simulation of the volatile behavior of the exchange rates were created models that treat the exchange rates as being prices of the financial assessments traded on efficient markets. The current exchange rate contains the currently available information and the changes observed reflect the effect of the new events that are unpredictable by definition.

The theory states that an accurate a priori prediction of the exchange rate evolution is impossible to be made but the subsequent explanation of the changes is possible. In order to eliminate these difficulties, the chaos theory and the nonlinear models are extensively used. The first researches have been carried out starting from 1980.

In the majority of situations these models are highly nonlinear and result in a wide range of dynamic behavior, including chaotic dynamics. There is a dispute over the manifestation of chaotic dynamics in exchange rates. There are many studies that are positive to the chaotic dynamics (Federici 2001, Westerhoff, Darvas 1998, Hommes 2005, Vandrocicz 2006) and also a number of studies that are rejecting the chaos in exchange rate (Brooks, Serletis).

The chaos theory demonstrates that even the simplest dynamical systems can exhibit at some point a very complex behavior. If the exchange rates variation is caused due to the chaotic nature of the system, this should lead to the fact that the smallest influences should have the effect of a nonlinearity over the exchange rates – exactly what happens in reality.

The first model presented demonstrates the fact that even the simplest models can exhibit chaotic behavior. [3]

The demand of foreign currency is determined as percentage of the deviation of current exchange rate towards the expected one.[2]

$$S_t = \alpha \left( \frac{e^e}{e_t} - 1 \right), \alpha \geq 0 \quad (2)$$

where
$e_t$ is the domestic price of the foreign currency
$e^e$ is the future estimated exchange rate
α is the sensitivity parameter

The trade balance ($T_i$) is a linear function depending on the current exchange rates and the corresponding exchange rate for the last period, written as deviation from the expected values and is given by the equation:

The expected exchange rate represents the stable state at which the speculators on the market do not wish to sell nor buy.

$$T_t = \beta(e_t - e^e) + \gamma(e_{t-1} - e^e) \quad \beta, \gamma > 0 \quad (3)$$

The clearing of the exchange markets writes as:

$$\Delta S_t = T_t \quad (4)$$

After replacing equations (2) and (1) in (4), equation (4) becomes:

$$\beta e_{t-1} e_t^2 - [(\beta + \gamma)e^* e_{t-1} - \gamma e_{t-1}^2 - ae^*]e_t - \alpha e^* e_{t-1} = 0 \quad (5)$$

The equation 5 has two roots, the positive one being considered for obvious reasons. The resulting nonlinear equation is:

$$e_t = \frac{[(\beta + \gamma)e^* e_{t-1} - \gamma e_{t-1}^2 - ae^*]}{2\beta e_{t-1}} + \frac{\sqrt{[(\beta + \gamma)e^* e_{t-1} - \gamma e_{t-1}^2 - ae^*]^2 + 4 * \beta e_{t-1} * \alpha * e_{t-1}}}{2\beta e_{t-1}} \quad (6)$$

for α=β=4 and γ=26.

The graphical representation of the solution $e_t$ show that the graph presents a peak value of 2.76 and a minimum value of 0.091. Any other value from outside the interval represented by these two values is attracted. The evolution of the system with the specified parameters is chaotic because satisfies the Ly-Yorke condition [3].

The Figure 2 illustrates the evolution of the system for two initial slightly different values: 0.2 and 0.2005 (the dotted line). The values of the two time series are identical for a short period of time (the first 10 iterations) and then the trajectories of the systems are diverging.

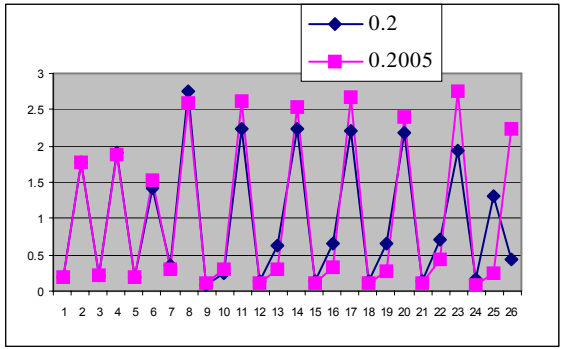

Fig. 2. The influence of the initial conditions.

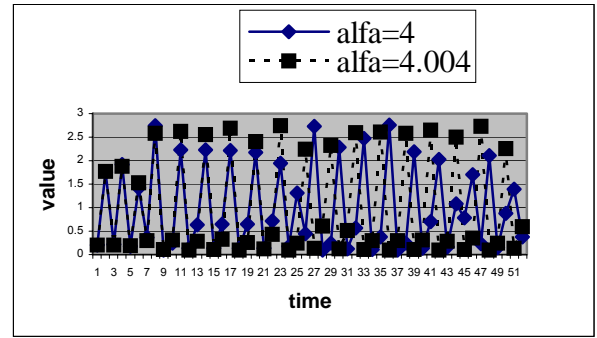

Fig. 2. The influence of the initial conditions.

The scatterplots for the two time series are provided to demonstrate the independence of the two time series after 10 iterations. The scatterplots presented in Figure 3 and Figure 4 one of the fingerprints of chaos: the distance between two trajectories starting from nearby points in the state space diverge over time.

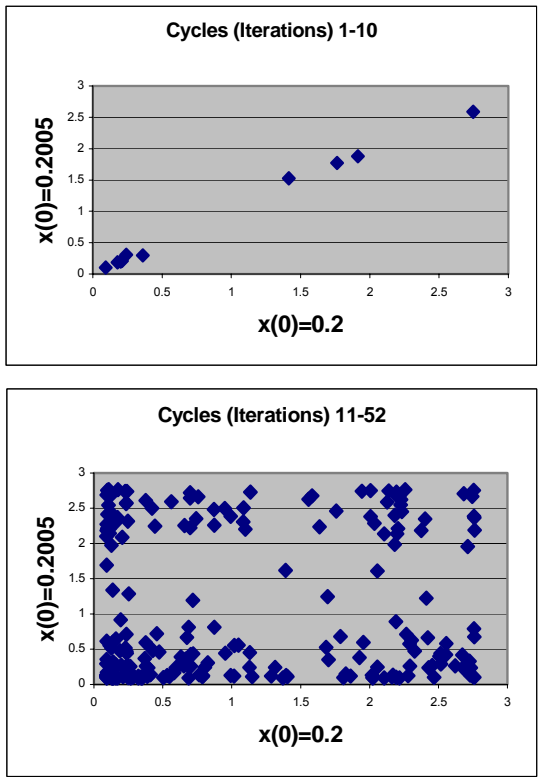

Fig. 1. The scatterplot for the first 10 iterations and the scatterplot for the last 41 iterations.

When the sensitivity parameter is varied, the same effects can be observed. Figure 4 presents the trajectories of the system for two very near values of α.

The apparently irrelevant changes can affect the longtime behavior of the exchange rate modeled using the Ellis model and some of these small shocks can determine the system to fall into the chaotic regime.

### 2.2 The model of the nonlinear feedback mechanism of the profit

The current spending of a firm can influence the value of the profit obtained at the end of the reference period. The profit will influence the spending over the next period. The dependence between the previous value of the profit and the current value is nonlinear because an increase of the spending does not reflect in an increase of the profit. The law of the decrease of the efficaciousness asserts that a certain mean value reaches minimum or maximum value when its magnitude equals the marginal value. One can invest in a certain production capability but this doesn't guarantee an unlimited increase of the production but the increase up to a certain point. Beyond that point the increase of the investment does not generates a corresponding increase of the production.

The dependence between the current profit and the previous profit can be modeled by using the equation:

$$\Pi_{t+1} = A\Pi_t - B\Pi_t^2 \quad (6)$$

The maximum profit $\Pi_{max}$ is supposed that it can be determined.

Dividing the equation (6) with $\Pi_{max}$ the following result is obtained:

$$\frac{\Pi_{t+1}}{\Pi_{max}} = A \frac{\Pi_t}{\Pi_{max}} - B\left(\frac{\Pi_t}{\Pi_{max}}\right)^2 \Pi^{max} \quad (7)$$

Let $\pi_t = \frac{\Pi_t}{\Pi^{max}}$ and the equation (7) becomes:

$$\pi_{t+1} = A\pi_t - B\pi_t \Pi^{max} \quad (8)$$

If we take $\Pi^{max} = \frac{A}{B}$ the equation above becomes the logistic equation:

$$\pi_{t+1} = A\pi_t - A\pi_t^2 = A(1 - \pi_t)\pi_t \quad (8)$$

The logistic map exhibits the same dependence on the initial condition: the slightest change of the initial condition causes a completely different evolution.

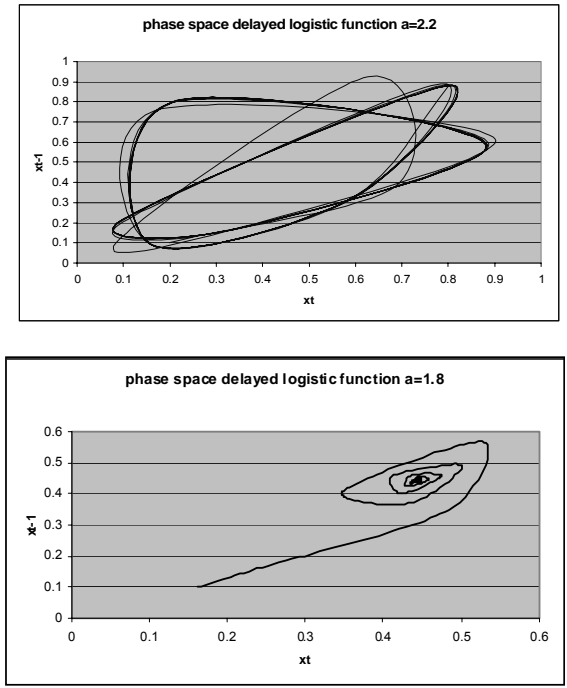

Fig. 5. Phase space portrait of the logistic delayed function $x_t = ax_{t-1}(1-x_{t-2})$. For small values of the parameter a the non-chaotic attractor is a point. For a larger value of the parameter the nonchaotic attractor is a limit cycle.

The complex behavior of the apparently simple functions can be observed using the bifurcation diagram. The bifurcation diagram (Figure 6) is an excellent tool allowing analyzing the behavior of a function by varying a control parameter (in the case of logistic function, the control parameter is A).

The logistic function is known to have a chaotic behavior with small isles of periodicity for a value of the parameter A greater that 3.57. For $A \in [3.57, 4]$ there are small areas of periodicity, the white stripes that can be observed in the figure. For A>4 the behavior is completely chaotic.

## 2.3 The K-Z model of Larrain

The theory and the models regarding the functioning of the capital markets initially developed on the hypothesis according to which these markets are efficient. The efficient-market hypotheses comprise a series of conditions which basically say that the prices of the assets and their turnover can be determined based on the supply and demand in the competitive market where there are rational agents. These rational agents quickly assimilate any piece of information that is relevant to determining the prices of the assets and their turnover, adjusting the price in accordance with this information. In other words, the agents do not have different comparative advantages in acquiring information.

That is to say that such a market does not provide opportunities to obtain a profit on an asset whose turnover is superior to the risk undertaken by the agent. Thus, the normal profits will be nil, taking into consideration the fact that the agents procure this piece of information and immediately incorporate it into the price of the assets. If the last piece of information and the current one are incorporated immediately into the price of the assets, then only a new piece of information, or however else we may call it - the "novelty", will be able to determine a change in prices.

As novelties are, by definition, unpredictable, then the changes in prices (or in turnover) will be unpredictable, too: no piece of information at time t or previous to this time will be able to help improve the forecast concerning the prices and turnovers (nor contribute to the decrease in the forecast errors made by the agents in this respect).

This forecast error independence towards the previous information is called the feature of orthogonality and is widely used in testing the efficient market hypothesis.

M. Larrain elaborated a model combining a classical description of a Keynesian economy with a non-linear model based on the evolution of the interest rates. Because he named the behavioural model "Z application" and the non-linear model "K application", this led to the so-called K-Z model. Larrain introduces the two components separately.

Thus, he observes that the future interest rates in the capital markets depend both on the previous interest rates (the technical analysis conception):

$$r_{t+1} = f(r_{t-n}), = 0,1,2,...n \quad (10)$$

where f is a non-linear function, and on a series of fundamental economic variables (fundamentalist conception):

$$r_{t+1} = g(Z) \quad (11)$$

Where $Z = (y, M, P, ...)$, y - being the real GNP, M - the money offer, P - the consumer price index etc.

The component (1a) shows that the future interest rate depends on its previous rates up to a certain lag n. This dependence of the future sizes on the previous ones is specific to the conception of the technical analysis of approaching the capital markets.

The exact form of $f(r_{t-m})$ is unknown, it may differ from one analyst to another.

Larrain chooses for this function the expression:

$$r_{t+1} = a + br_t^n - cr_t^{n+1}$$

where we can notice that if c=b becomes the logistic equation, which is known to have a chaotic behaviour for certain values of the control parameter b (or c):

$$r_{t+1} = a - br_t^n(1 - r_t) \quad (12)$$

This represents the K component (application) of the model.

In what follows we shall present the way in which is built the Z component:

$$r_{t+1} = dy_t + eP_t - f \cdot M_t - g\sum(Y_t - c_t) \quad (13)$$

where d, e, f, g are constants and $y_t$ represents the real GNP, $M_t$ – the money offer (expressed through the aggregate $M_t$), $P_t$ the - the consumer price index, $Y_t$ the real personal and $c_t$ – the real personal consumption.

This component reflects the fundamentalist conception according to which the interest rates in the capital markets depend on the evolution of fundamental sizes.

Larrain combines the two components, K and Z, in one single expression as follows:

$$r_{t+1} = a - br_t^n(1-r_t) + dy_t + eP_t - f \cdot M_t - g\sum_t (Y_t - c_t) \quad (14)$$

This expression shows that the future interest rates are a combined function of technical and fundamental factors. Meanwhile, the former or latter of the two components can dominate the other one. Thus, during the stability periods, the capital markets are efficient and the interest rates will depend on the Z component, to a larger extent. In the unstable periods of the markets in question, the investors lose their trust in fundamental variables, making decisions by extrapolating tendencies. Thus, it is the K component that becomes dominant.

In this situation, under certain circumstances, the c control parameter can take values in intervals for which the logistic equation has a chaotic behaviour, thus inducing crises and chaos episodes in the markets in question.

The tests made with Larrain's model led to a series of interesting conclusions.

Thus, for the stable capital markets, such as the bond market or the security market, the obtained forecasts covered quite well the evolution of the interest rates observed in reality, which for such a model represents a success. Still, for the estimation of the equation parameters (14), the model used techniques of linear regression, which annulled the premise that one or another of the two components can be dominant in one period or another.

In order to introduce such an alternation of the dominance of the K-Z application components it is necessary that the parameters of the function in question should be variable in time, which the model in its initial form cannot allow.

Improving such a model could reconcile the two big tendencies in the analysis of the capital markets, technical and fundamentalist, offering a powerful instrument of forecasting these markets.

## 3 CONCLUSION

Chaos is can be found almost everywhere in the nature. Chaos theory and fractals are currently applied in the study of the natural phenomenon.

An essential condition needed in order that chaos to emerge is to have nonlinear systems. In fact very few of all models are purely linear, the vast majority of the systems are nonlinear.

The paper emphasizes two of the features of the chaotic systems: dependence to initial conditions and the divergence of nearby trajectories.

The chaos theory has a significant impact on economy and especially on capital markets. If the behavior of one economic system is proved to be chaotic this guarantees that, using appropriate methods, a short-term prediction can be made.

**Sorin Vlad** graduated from "Ştefan cel Mare" University of Suceava, Electrical Engineering Faculty, section Computer and System's Science, 1998. He is Phd Candidate in Computer Science, field of research - Chaotic systems behavior modeling, University "Ştefan cel Mare" of Suceava, since 2004. He is with the Informatics Department, University "Ştefan cel Mare" of Suceava, Economic Sciences and Public Administration. His research interests include: neural networks, expert systems, logic programming, chaos theory, chaotic time series analysis and prediction.

**Paul Pascu** graduated from "Gheorghe Asachi" University of Iasi, Faculty of Computer Science, 1999. He is Phd Candidate in economie, field of research – Cybernetics models in economy, Economics Bucharest Academy of Economic, since 2006. He is with the Informatics Department, University "Ştefan cel Mare" of Suceava, Economic Sciences and Public Administration. His research interests include: cybernetic economy, economic modeling, databases, logic programming.

**Nicolae Morariu** graduated from "Alexandru Ioan Cuza" University of Iaşi, Mathematics - Mechanics Faculty, section Computing Machines, 1972. He obtained the PhD degree from "Ştefan cel Mare" University of Suceava in 2004, with the thesis entitled: *Contribution to the development of data and knowledge bases*. His postgraduate activity includes: The design and implementation of the applications and informatics systems within the Regional Electronic Computing Center of Suceava and Informatics Services Society of Suceava (1972-1993), research projects within the national research programs (1993-2002), SSI Suceava design-research manager department (1998-2002), associate professor "Stefan cel Mare" University of Suceava, Electrical Engineering Faculty (1991-1998). Presently he is associate professor at the Economic Sciences and Public Administration Faculty, "Stefan cel Mare" University of Suceava. His research interests include databases: FoxPro, Access, Oracle administration and SQL programming, Deductive databases. Artificial intelligence: expert systems, pattern recognition, neural networks, vegetal infogenetics.